\documentclass[aps,comment,a4,superscriptaddress,preprintnumbers,twoside,twocolumn,floatfix]{revtex4}

\usepackage{bm}
\usepackage{graphicx}
\usepackage{amsfonts}
\usepackage{amsmath}
\usepackage{amssymb}
\usepackage{amsthm}
\usepackage{bbm}
\usepackage{setspace}
\usepackage{helvet}
\usepackage{times}

\theoremstyle{plain}

\begin{document}

\title{Comment on ``Feshbach-Einstein condensates" by V. G. Rousseau and P. J. H. Denteneer}

\author{Mar\'ia Eckholt}
\affiliation{Max-Planck-Institut f\"ur Quantenoptik, Hans-Kopfermann-Stra\ss e 1, 85748 Garching, Germany}
\author{Tommaso Roscilde}
\affiliation{Laboratoire de Physique - ENS Lyon, 46 All\'ee d'Italie, 69007 Lyon, France}
\date{\today}

\begin{abstract}
\end{abstract}



\maketitle

 In a recent letter \cite{RousseauD09}, making use of quantum Monte Carlo (QMC)
 simulations Rousseau and Denteneer (RD) have 
 investigated a two-species Bose-Hubbard (BH) model in a one-dimensional optical lattice, 
 with coherent conversion between two particles of a species and one of the other, 
 mimicking the atom-molecule coherence occurring in cold gases in optical 
 lattices close to a Feshbach
 resonance \cite{Thalhammeretal06, Syassenetal07}. In the notation of Ref.~\onlinecite{RousseauD09}, 
 the Hamiltonian under investigation is 
 \begin{eqnarray}
 &&\null\hspace{-.5cm}{\cal H} = -t_a \sum_i \left( a_i^{\dagger} a_{i+1} + \text{h.c.} \right) 
 - t_m \sum_i \left( m_i^{\dagger} m_{i+1} + \text{h.c.} \right) \nonumber \\
 &&\null\hspace{-.7cm}+ U_{aa} \sum_i n_i^{(a)} \left(n_i^{(a)}-1\right) + U_{mm} \sum_i n_i^{(m)} \left(n_i^{(m)}-1\right) \nonumber \\
 &&\null\hspace{-.7cm}+ U_{am} \sum_i n_i^{(a)} n_i^{(m)} 
 + D \sum_i n_i^{(m)} + g \sum_i \left( m^{\dagger}_i a_i^2 + \text{h.c.}\right)~~~
 \label{e.ham}
 \end{eqnarray}
 where $t_a$ and $t_m$ are the atomic and molecular hoppings, $U_{aa}$ and $U_{mm}$ 
 are the interatomic and intermolecular interactions, 
 $U_{am}$ is the atom-molecule interaction, 
 $D$ is the ``detuning" between atomic and molecular states (the true 
 detuning being $D-U_{aa}$) and $g$ is the amplitude of atom-molecule
 conversion. This model was previously investigated by the same
 authors in Ref.~\onlinecite{RousseauD08} for a large variety of 
 parameters. 

 One of the main claims of Ref.~\onlinecite{RousseauD09} is the existence of a 
 novel phase, dubbed \emph{super Mott} (SM), which is characterized by 
 zero compressibility (namely absence of fluctuations in the total 
 particle number $N = N_a + 2 N_m$) and finite superfluid fraction of both
 the atoms ($\rho_s^a$) and the molecules ($\rho_s^m$).
 Moreover, the superflows of atoms and molecules appear to be anticorrelated, 
 namely the correlated superfluid density ($\rho_s^{\rm cor}$) turns out to be zero. 
 In this comment we argue that the claimed SM phase does \emph{not} contain
 any superfluid component, and that the individual superfluid densities
 of atoms and molecules $\rho_s^a$ and $\rho_s^m$ are not well 
 defined in the model of Eq.~\eqref{e.ham}. In fact, the only meaningful
 superfluid density is the correlated one $\rho_s^{\rm cor},$ 
 which turns out to be vanishing, revealing a normal phase. We corroborate
 the above statement with the explicit numerical calculation of the correlation function 
 associated with the coherent counterflow of atoms and molecules, and we
 show that this correlator is short-ranged in the supposed SM region. 

 \begin{figure}[h!]
 \includegraphics[width=60mm]{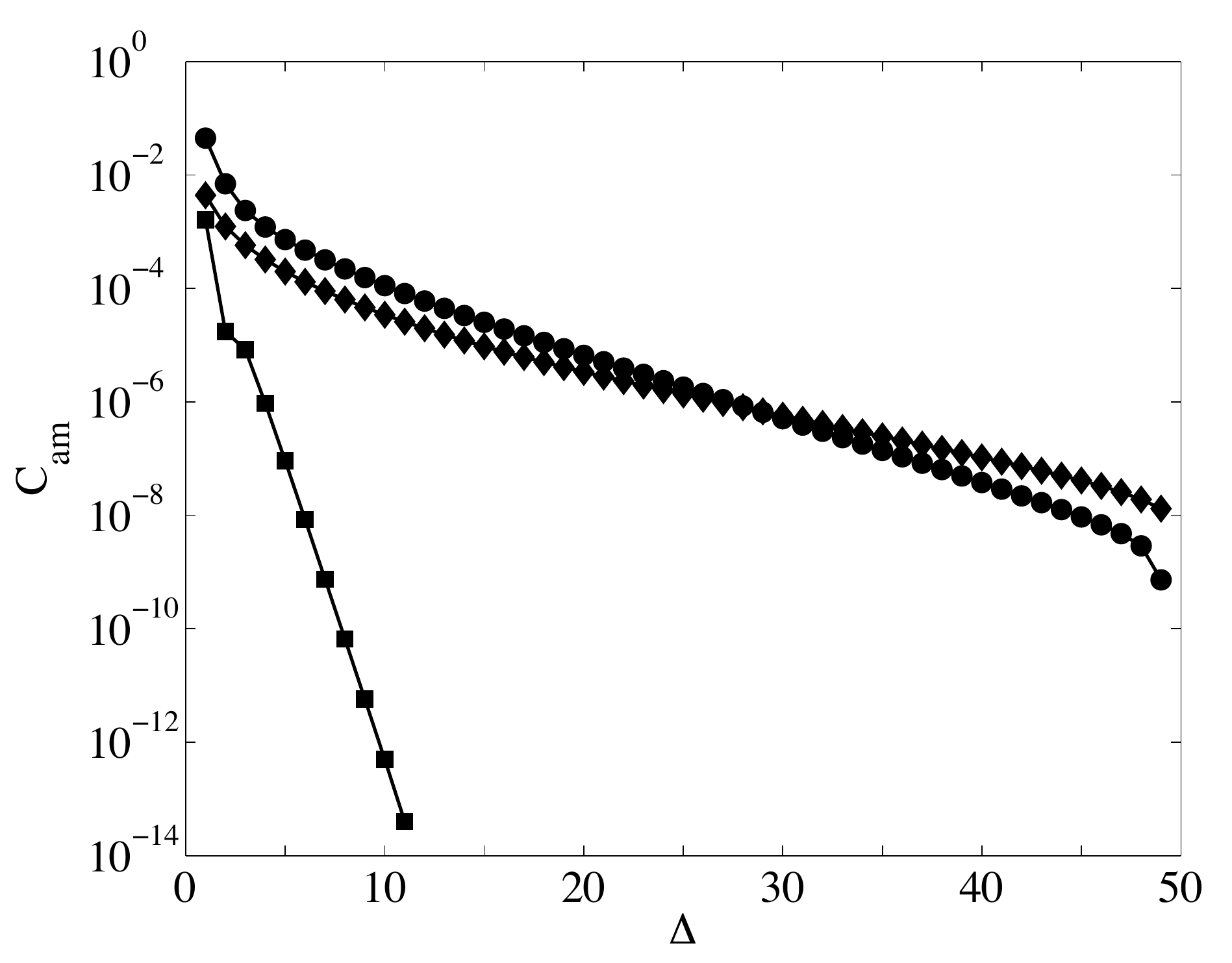}
 \caption{Atom-molecule correlator for a $L=100$ chain with  
 parameters:  $(D/t_a, \mu/t_a) = (3, 8.6)$ [squares], 
 (7, 9.6) [circles], (13, 10.6) [diamonds].
 The other parameters are $U_a = 8t_a$, $U_{am} = 12 t_a$,
 $U_m = \infty$, $g = 0.5 t_a$, $t_m = 0.5 t_a$.
 According to 
Fig.~10 of Ref.~\onlinecite{RousseauD08}
 all three parameter sets should give a SM phase .}
 \label{f.Cam}
 \end{figure}

 At $T=0$ the superfluid density of one bosonic species is defined via the
 energy cost of a boost in the phase of the operators associated
 to that species. E.g. for the atoms, considering the 
 phase shift $a_j \to a_j e^{i\phi j},$ one defines 
 $\rho_s^a  = \frac{1}{2L} \partial^2 E_0/\partial \phi^2 |_{\phi=0}$ (where $E_0$ is 
 the ground state energy and $L$ is the size of the system). For a single
 species BH model, the explicit calculation of the superfluid
 density within the path-integral formalism shows that this quantity 
 can be estimated via QMC simulations on a system
 with periodic boundary conditions by means
 of the fluctuations of the winding number ($W$) of the worldlines associated
 with the motion of particles in imaginary time \cite{PollockC87, Cuccolietal03}. 
 When particle-number conservation holds --- as in the single-species BH model --- 
 supercurrents are topological invariants (independent of the 
 connected surface through which the current is measured), and indeed so 
 is the winding number for each configuration sampled by the Monte Carlo.
 Yet in the case of Eq.~\eqref{e.ham}, the atom and molecule numbers are
 not conserved separately, so that neither the currents of each species are
 not topological invariants nor the atomic and 
 molecular winding numbers, $W_a$ and $W_m$. The only conserved
 quantity is the total number $N,$ and hence the only meaningful superfluid
 density is the one associated with the correlated superflow of atoms 
 and molecules, captured by the correlated winding number $W_{\rm cor} = W_a + 2 W_m$
 introduced in Refs.~\onlinecite{RousseauD09,RousseauD08}. 
 From a technical point of view, this corresponds to the 
 fact that an arbitrary phase boost  of the operators, 
 $a_j \to a_j e^{i\phi_a j},$ $m_j \to m_j e^{i\phi_m j},$ produces a 
 rotation of the atom-molecule conversion term
 $m^{\dagger}_j a_j^2 \to m^{\dagger}_j a_j^2 e^{i(2\phi_a-\phi_m)j}$
 which grows with the position $j$ (hence leading to an energy variation 
 in the ground state which is not infinitesimal in the limit $\phi_{a,m} \to 0$),
 unless $\phi_m = 2\phi_a,$ which corresponds to probing
 the correlated response of atoms and molecules, namely 
 the correlated superfluid density. In fact, following  Refs.~\onlinecite{PollockC87, Cuccolietal03}
 it is straightforward to prove that the correlated superfluid density, defined
 in Ref.~\onlinecite{RousseauD09} as
 $\rho_s^{\rm cor} = \lim_{T\to 0} (k_B T L/2)  \langle W_{\rm cor}^2 \rangle$,
 can be written as $\rho_s^{\rm cor} = \frac{1}{2L} \partial^2 E_0/\partial {\phi_a}^2 |_{\phi_a=0}$
 with the correlated boost $\phi_m = 2\phi_a$. 
 Hence the only meaningful superfluid density is the correlated one; which, 
 as shown in Refs.~\onlinecite{RousseauD09, RousseauD08}, is perfectly
 vanishing in the SM phase. Consequently the supposed SM phase is 
 actually \emph{normal}, and its name is misleading.
 
  As argued in Refs.~\onlinecite{RousseauD09, RousseauD08}, the superfluid
 nature of the SM phase stems from counterflow of pairs of atoms and molecules, 
 leading to a zero net mass current. Counterflow superfluidity (CSF) has been 
 theoretically predicted in repulsive binary mixtures of bosons (with particle number
 conservation of both species) and it is associated to condensation (or
 quasi-condensation in one dimension) of 
 composite objects made of one particle of a given species and one hole
 of the other species \cite{KuklovS03,Huetal09}. In analogy to CSF, if the
 SM phase contained counterflowing superfluid components, one would then
 expect  composite objects made of two atoms and a molecular hole --- or of two atomic
 holes and one molecule --- to quasi-condense. We have explicitly checked
 this by investigating the atom-molecule correlation function 
 $C_{am}(\Delta) = \langle (a^{\dagger}_{L/2})^2m_{L/2} 
 m_{L/2+\Delta}^{\dagger} (a_{L/2+\Delta})^2\rangle
 - \langle (a^{\dagger}_{L/2})^2m_{L/2}\rangle  
 \langle m_{L/2+\Delta}^{\dagger} (a_{L/2+\Delta})^2\rangle$    
 for parameters which are claimed by RD to correspond to 
 a SM phase. Fig.~\ref{f.Cam} shows the above quantity calculated numerically
 by a variational Matrix Product State Ansatz \cite{Verstraeteetal04} 
 with bond dimension $D=60$, for a chain of length $L=100$
 with open boundary conditions;
 all the parameters are chosen so as to be in the SM phase (according to  
 Fig.~10 of Ref.~\onlinecite{RousseauD08}) and close to the resonance
condition $D = U_a$. We find that the $C_{am}$ correlator decays 
\emph{exponentially} for all the investigated parameters, as well
as all the correlators to lower order. Hence no quasi-condensation phenomenon
is observed in the supposed SM phase, confirming its fully normal character. 

We acknowledge fruitful discussions with J.-J. Garc\'ia-Ripoll and S. Capponi, 
and useful correspondence with V. G. Rousseau and P. J. H. Denteneer.

\end{document}